\documentclass[doublecol]{epl2}
\usepackage[utf8x]{inputenc}
\usepackage{amsmath,amssymb,graphicx}
\newcommand{\vek}[1]{\boldsymbol{#1}}
\newcommand{\abs}[1]{\lvert #1\rvert}
\title{Influence, originality and similarity in directed acyclic graphs}
\author{S. Gualdi\inst{1} \and M. Medo\inst{1} \and Y.-C. Zhang\inst{1,2}}
\shortauthor{S. Gualdi \etal}
\institute{
  \inst{1} Physics Department, University of Fribourg, CH-1700 Fribourg, Switzerland\\
  \inst{2} Web Sciences Center, School of Computer Science and
  Engineering, University of Electronic Science and Technology of
  China, Chengdu 610054, P. R. China}
\abstract{We introduce a framework for network analysis based on
random walks on directed acyclic graphs where the probability
of passing through a given node is the key ingredient. We
illustrate its use in evaluating the mutual influence of nodes
and discovering seminal papers in a citation network. We further
introduce a new similarity metric and test it in a simple
personalized recommendation process. This metric's  performance
is comparable to that of classical similarity metrics, thus
further supporting the validity of our framework.}
\begin{document}

\maketitle

The past two decades have witnessed a network
revolution~\cite{Newman03} fueled by the ever-increasing
computer computational power at our disposal and by the
availability of rich datasets mapping virtually all fields of
human activity~\cite{DJW04,GoBa07}. Complex networks and
algorithms based on these resources found their application in
the most diverse fields, ranging from nonlinear dynamics and
critical phenomena~\cite{Strogatz2001,Dorogo2008} to social and
economic systems~\cite{Jackson2008}. Random walks are among the
most prominent classes of processes taking place on networks,
being employed in importance rankings for the World Wide
Web~\cite{Langville2006}, recommender systems~\cite{Zhou2010},
disease transmission models~\cite{Altmann93}, nodes
similarity~\cite{Fouss07} and many other areas~\cite{Young01}.

A relatively less-studied class of networks is represented by
directed acyclic graphs (DAGs) which occur in both natural and
artificial systems. Their acyclicity (absence of directed
cycles) stems either from an implicit time ordering (as in
citation networks where only past papers can be cited) or from
natural constraints (as in food webs). Even when nodes of a DAG
do not have time stamps attached, a causal structure with all
edges pointing from later to earlier nodes can always be
recovered. Theoretical models exist for building random DAGs
with fixed degree sequences or with fixed expected
degrees~\cite{Karrer09a,Karrer09b}.

Acyclicity turns out to be highly advantageous to filter
information through a random walk process. If we consider a
random walk on a generic network, the probability of passing
through a given node---which we refer to as passage
probability---is usually not a meaningful quantity as it may
well be equal to one for all nodes in the network. The situation
is rather the opposite if we instead consider a DAG, as every
random walk along the network's edges comes to an end when a
root node with zero out-degree is reached.

In this Letter we introduce an analytical framework for DAGs to
quantify the influence of one node over another based on the
passage probability and discuss its applications. In particular
we propose a method to identify papers fundamental to the growth
of a given research area and define a new similarity metric.
Relation to PageRank, which has been used to citation data
before~\cite{Chen07} (see~\cite{Franceschet11} for a historical
perspective of PageRank and other fields of its applicability),
is also discussed. We test our framework on citation
data provided by the American Physical Society and we show that:
i) the proposed method is able to uncover seminal papers even if
they do not have particularly high citation counts, (ii) the
similarity metric performs well when used as a component of a
simple recommendation algorithm~\cite{AdoTuzi05}. Note that the
time dimension, neglected by many information filtering techniques,
is implicitly taken into account by acting on a DAG. While we use
academic citation data to test our model and often refer to papers
and citations instead of nodes and edges, majority of this work is
general and applicable to other DAGs such as those representing
family trees and reference networks of patents~\cite{JaTra02} and
legal cases~\cite{Fow07}.

Consider a directed acyclic graph composed of $N$ nodes and $L$
directed edges pointing from newer to older nodes. In- and
out-degree of node $x$ are denoted as $k_x^{in}$ and
$k_x^{out}$, respectively. We further denote by $\mathcal{A}_x$
the set of nodes that can be reached from node $x$ ($x$'s
ancestors) and by $\mathcal{P}_x$ the set of nodes from which
$x$ can be reached ($x$'s progeny). Since the network is
acyclic, $\forall x:\ \mathcal{A}_x\cap\mathcal{P}_x=\emptyset$.
A random walk starting in node $x$ can be encoded in an
$N$-dimensional vector $\vek{G}_x$ whose $i$th component
represents the probability of passing through node $i$ (see
Fig.~\ref{fig:illustr}a for an illustration). Thanks to the
network's acyclicity, $\vek{G}_x$ fulfills the equation
\begin{equation}
\label{trans}
\vek{G}_x=\mathsf{W}\vek{G}_x
\end{equation}
where $\mathsf{W}$ is the transition matrix with elements
$W_{ij}=1/k_i^{out}$ if $i$ cites $j$ and $W_{ij}=0$ otherwise.
The boundary condition for Eq.~(\ref{trans}) is given by
$(\vek{G}_x)_x=1$ which reflects that any random walk certainly
passes through its starting point. (One can also obtain $\vek{G}_x$
by simply following the random walk starting at node $x$ as it is
done in Fig.~\ref{fig:illustr}a.) Elements of $\vek{G}_x$ are by
definition positive for all nodes in $\mathcal{A}_x$ and zero for
all other nodes. Nodes without out-going links are represented by
a zero column in $\mathsf{W}$ and act as sinks for the random walk.

\begin{figure}
\centering
\includegraphics[scale=0.9]{1a}\hspace{40pt}
\includegraphics[scale=0.9]{1b}
\caption{Comparison of a random walk starting at $X$ (a) with
passing of ``genes'' (b). According to the description in the
main text, $\vek{c}_i=\vek{e}_i$ for $i=1,2,3$,
$\vek{c}_4=\tfrac13(\vek{c}_1+\vek{c}_2+\vek{c}_3)+
\vek{e}_4$, $\vek{c}_5=\vek{c}_3+\vek{e}_5$, $\vek{c}_6=
\tfrac13(\vek{c}_4+\vek{c}_5+\vek{c}_3)+\vek{e}_6=
\tfrac19\vek{e}_1+\tfrac19\vek{e}_2+\tfrac79\vek{e}_3+
\tfrac13\vek{e}_4+\tfrac13\vek{e}_5+\vek{e}_6$. Coefficients in
$\vek{c}_6$ agree with the corresponding passing probabilities
in (a). Note that while the random walk proceeds from top to
bottom, genetic composition propagates from bottom to top.}
\label{fig:illustr}
\end{figure}

To obtain a compact formalism, we construct an $N\times N$
matrix $\mathsf{G}$ where column $x$ is equal to $\vek{G}_x$.
Elements of this matrix have simple interpretation: $G_{yx}$
represents the probability of passing through node $y$ when
starting in node $x$. One may check that
$G_{yx}=\sum_{n=0}^{\infty}(\mathsf{W}^{n})_{yx}$ (since
$\mathsf{W}$ is a transition matrix, $(\mathsf{W}^n)_{yx}$ is
the probability of moving from $x$ to $y$ over a path of length $n$).
Note that while Eq.~(\ref{trans}) reminds an equation for stationary
occupation probabilities, this not the case: Unlike the classical
random walk utilized by PageRank, the stationary occupation probability
here is zero for all nodes due to the presence of sinks (the
relation between our framework and PageRank is discussed in
detail below). This concept can be readily generalized for a
weighted DAG by assuming that the probability of choosing an
outgoing edge is proportional to the edge's weight.

It is instructive to complement the above random walk approach
with an analogy based on genes spreading in a population. In the
context of citation data, consider vectors of ``genetic''
composition of papers and assume that each paper's vector is
obtained by averaging the vectors of the cited papers (inherited
knowledge) and by adding the paper's contribution (new
knowledge). A similar model based on genetic composition of
scientific papers has been shown to reproduce many quantitative
features of science~\cite{Gilbert97}. Fig.~\ref{fig:illustr}
illustrates this process on a toy network. For example,
$\vek{c}_6=\tfrac13(\vek{c}_1+\vek{c}_4+\vek{c}_5)+\vek{e}_6$
where $\vek{e}_6$ represents contribution of paper $6$ which is,
by definition, orthogonal to contribution vectors of all
previous papers. Vectors $\vek{e}_1,\vek{e}_2,\dots$ therefore
constitute a basis of a space of growing dimension. The
accumulation of knowledge is reflected in the lack of
normalization of the composition vectors $\vek{c}_x$ which are
of greater magnitude for recent papers than for old ones.
From a correspondence between all possible paths from $x$ to $y$
and possible ways how composition $\vek{c}_y$ can propagate to $x$,
it is straightforward to show that when composition of a paper is
written in terms of the base vectors, coefficients of respective
base vectors are equal to the passage probabilities obtained by the
random walk approach and hence $\vek{c}_x=\vek{G}_x$ (see
Fig.~\ref{fig:illustr}). We can say that the previously introduced
passage probabilities $\vek{G}_x$ represent influence of past
papers on paper $x$ and, at the same time, ``genetic'' composition
of paper $x$.

Given our understanding that $G_{xy}$ quantifies the influence
of $x$ on $y$, we may introduce the total aggregate impact of
node $x$
\begin{equation}
\label{flux_genes}
I_x=\sum_y G_{xy}.
\end{equation}
where the number of non-zero terms in the summation is
$P_x:=\abs{\mathcal{P}_x}$ (which we refer to as the progeny
size of node $x$). The value $I_x$ is not meaningful by itself
because it is naturally biased by the size of $\mathcal{P}_x$.
This makes it sensitive to the time of the paper's appearance
(old nodes tend to have greater progenies) and to the amount of
literature in this paper's research field. It is therefore more
informative to plot $I_x$ vs $P_x$. A large value of $I_x/P_x$
is achieved when the influence of $x$ is effectively channeled
to the papers in $\mathcal{P}_x$: for example when even papers
that do not cite $x$ directly refer mostly to papers citing $x$.
Therefore we expect outliers in the plane $(P_x,I_x)$ to be
seminal papers which founded new branches of research.

It is illustrative to discuss the relation between the aggregate
impact $I_x$ and the Google PageRank score. To do that, we
combine Eqs.~(\ref{trans}) and (\ref{flux_genes}) to write $I_x$
as a solution of the self-consistent equation
\begin{equation}
\label{Ix-iter}
I_x = 1 + \sum_y W_{yx}I_y
\end{equation}
where $I_x:=1$ for all nodes without progeny (\emph{i.e.},
$k_x^{in}=0$). The structure of this equation resembles that of 
the classical PageRank equation. The similarity can be enhanced
further if instead of the ``gene'' composition spreading
discussed above, we consider its normalized version. This
normalized spreading is achieved by assuming that each paper's
genetic vector is composed by a fraction $(1-\alpha)$ of its
original contribution plus a fraction $\alpha$ of the average
over its parents' genetic vectors (thus the vector's norm is
fixed to one for all papers with at least one ancestor). Hence
we obtain a new matrix of genetic composition,
$\mathsf{G}^\alpha$ which in turn can be used to compute new
aggregate impact $I^\alpha_x$. The self-consistent equation for
$I^\alpha_x$ now has the form
\begin{equation}
\label{Ixa}
I^\alpha_x = 1-\alpha + \alpha\sum_y W_{yx}I^\alpha_y
\end{equation}
where $I^\alpha_x:=1-\alpha$ for all nodes without progeny. Up
to replacing $1-\alpha$ with $(1-\alpha)/N$ (which only affects
the overall scale of $I^\alpha_x$), this equation is identical to
the equation of the PageRank: $\alpha$ and $1-\alpha$ are the
probabilities that the random walk follows an
existing link and jumps, respectively, and $I^\alpha_x$ is the
PageRank value of node $x$. Since the term $1-\alpha$
only sets the scale of $I^\alpha_x$ and in the limit
$\alpha\to1$ the propagation term $\alpha\sum_y W_{yx}I^\alpha_y$
in Eq.~(\ref{Ixa}) is equal to that in Eq.~(\ref{Ix-iter}), we
see that rankings of nodes according to the aggregate impact
$I_x$ and the limit PageRank value $\lim_{\alpha\to1} I^\alpha_x$
are equivalent.

Both $I_x$ and $I^\alpha_x$ are naturally biased by the progeny
size of node $x$. In the case of $I^\alpha_x$, this bias can be
partially removed by setting $\alpha<1$ which leads to impact
spreading mainly over a local neighborhood. In the case of
$I_x$, we remove the bias by placing the nodes in the plane
$(I_x,P_x)$ which allows us to better distinguish exceptional
nodes than the one-dimensional PageRank value with one parameter
($\alpha$). While PageRank certainly has its merit for the WWW,
in what follows we attempt to show that influence and impact
propagating without damping are useful for DAGs.

\begin{figure}
\centering
\includegraphics[scale=0.28]{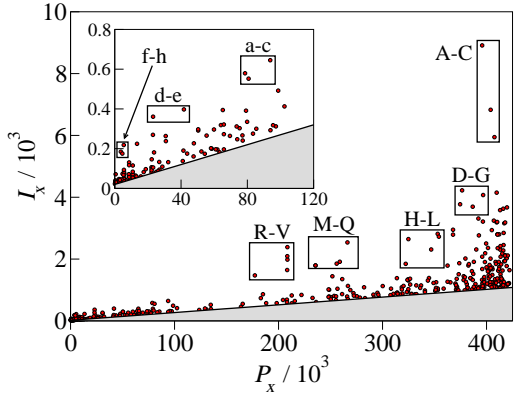}
\caption{Total influence of papers $I_x$ versus their progeny
size $P_x$ for the APS citation data (for clarity, only $413$
papers with $I_x>20+P_x/400$ are shown). Details about the
marked outliers are given in Tab.~\ref{tab:seminal}.}
\label{fig:seminal}
\end{figure}

We now illustrate our ideas on the citation data provided by the
American Physical Society (APS). This data contains all 449\,705
papers published by the APS from 1893 to 2009 together with
their citations to the APS journals. To make the data strictly
acyclic, we do not consider a small number of citations that are
between papers of the same print date; we are then left with
4\,672\,812 citations. Fig.~\ref{fig:seminal} shows all papers
published by the APS after 1940 and reveals an expected linear
relationship between $I_x$ and $P_x$ with several outstanding
papers whose influence is much greater than that of other papers
of the same progeny size. (Papers published before 1940 are
omitted because of the data sparseness which is amplified by
the limitation of our data to citations to and from the APS
journals.) Table~\ref{tab:seminal} lists the outliers together
with scientific prizes as a proxy for their quality. While our
results are affected by using only the APS
citations\footnote{For example, paper P which is not (to the
best of our knowledge) particularly outstanding owes its high
total impact to the fact that it is the only paper in the APS
data cited by the high-impact paper Q. Since paper Q in reality
cites many more papers, paper P probably wouldn't excel if
complete citation data would be used for the analysis (this has
been already discussed in~\cite{Chen07}). Similar problems arise
for those research fields where the original work was not
published on APS journals (take high-temperature
superconductivity, for example).}, one can conclude that
majority of these outlying papers really represents exceptional
research. While it is not our goal to rank the papers, one could
achieve that for example by dividing $I_x$ by the average $I_x$
of papers with the same progeny size $P_x$, thus making papers
of different age comparable.

\begin{table*}[t]
\centering
{\footnotesize
\begin{tabular}{cllrcrr}
\hline
\textbf{Id} & \textbf{Title} & \textbf{Authors} & \textbf{Year} & \textbf{Prize}& \textbf{PR} & \textbf{CR}\\
\hline
A  & Statistics of the Two-Dimensional Ferromagnet\dots & H. A. Kramers, G.H. Wannier & 1941 & LM & 54 & 1\,645\\
B  & Crystal Statistics in a Two-Dimensional Model\dots & L. Onsager & 1944 & NP & 8 & 87\\
C  & Theory of Superconductivity & J. Bardeen, \textit{et al.} & 1957 & NP & 2 & 10\\
D  & The Maser--New Type of Microwave Amplifier,\dots & J. Gordon, \textit{et al.} & 1955 & NP & 369 & 14\,517\\
E  & Infrared and Optical Masers & A. Schawlow, C. Townes & 1958 & NP & 171 & 2\,108\\
F  & Population Inversion and Continuous Optical Maser & A. Javan \textit{et al.} & 1961 & + & 169 & 14\,517\\
G  & Dynamical Model of Elementary Particles Based on\dots & Y. Nambu, G. Jona-Lasinio & 1961 & NP & 24 & 50\\
H & Self-Consistent Equations Including Exchange and\dots & W. Kohn, L. Sham & 1965 & NP & 1 & 1\\
I & Inhomogeneous Electron Gas & P. Hohenberg, W. Kohn & 1964 & MPM & 3 & 2\\
J & A Model of Leptons & S. Weinberg & 1967 & NP & 6 & 18\\
K & Static Phenomena Near Critical Points:\dots & L. Kadanoff, \textit{et al.} & 1967 & MPM & 58 & 355\\
L & Radiative Corrections as the Origin of Spontaneous\dots & S. Coleman, E. Weinberg & 1973 & DM & 31 & 75 \\
M & Scaling Theory of Localization:\dots & E. Abrahams, \textit{et al.} & 1979 & NP & 11 & 24\\
N & New Measurement of the Proton Gyromagnetic Ratio\dots & E.R. Williams, P.T. Olsen & 1979 & & 150 & 26\,327\\
O & New Method for High-Accuracy Determination of\dots & K. Klitzing & 1980 & NP & 32 & 134\\
P & Cluster Formation in Two-Dimensional Random Walk & H. Rosenstock, C. Marquardt & 1980 &  & 109 & 217\,150\\
Q & Diffusion-Limited Aggregation\dots & T.A. Witten, L.M. Sander & 1981 & + & 17 & 64\\
R & Electronic Structure of BaPb$_{1-X}$Bi$_{X}$O$_{3}$ & L.F. Mattheiss, D.R. Hamann & 1983 &  & 106 & 4\,224\\
S & Bulk Superconductivity at 36 K in La$_{1.8}$Sr$_{0.2}$CuO$_4$ & R.J. Cava \textit{et al.} & 1987 &  & 37 & 1\,086\\
T & Evidence for Superconductivity above 40\,K In\dots & C.W. Chu \textit{et al.} & 1987 & & 40 & 606\\
U & Superconductivity at 93 K in a New Mixed-Phase\dots & M.K. Wu \textit{et al.} & 1987 & + & 19 & 102\\
V & Self-Organized Criticality: An Explanation of\dots & P. Bak \textit{et al.} & 1987 & + & 16 & 47 \\
\textbf{a} & Teleporting an Unknown Quantum State via\dots & C.H. Bennett \textit{et al.} & 1993 & + & 53 & 26 \\
\textbf{b} & Bose-Einstein Condensation in a Gas of Sodium Atoms & K.B. Davis \textit{et al.} & 1995 & NP & 63 & 27 \\
\textbf{c} & Evidence of Bose-Einstein Condensation in\dots & C.C. Bradley \textit{et al.} & 1995 & + & 99 & 51 \\
d & TeV Scale Superstring and Extra Dimensions & G. Shiu, S.-H.H. Tye & 1998 & & 216 & 3\,991 \\
\textbf{e} & Small-World Networks: Evidence for a Crossover Picture & M. Barthélémy, L.A.N. Amaral & 1999 & + & 658 & 9\,872 \\
\textbf{f} & Negative Refraction Makes a Perfect Lens & J.B. Pendry & 2000 & DM & 279 & 192 \\
\textbf{g} & Composite Medium with Simultaneously Negative\dots & D.R. Smith \textit{et al.} & 2000 & + & 433 & 459 \\
\textbf{h} & Statistical Mechanics of Complex Networks & R. Albert, A.-L. Barabási & 2002 & VNM & 112 & 59 \\
\hline
\end{tabular}}
\caption{An approximately time-ordered list of the papers marked
in Fig.~\ref{fig:seminal} (labels agree with those marked in the
figure). To evaluate the quality of the list, we indicate the
most important prize received by the authors for research
pertinent to the listed papers (LM=Lorentz Medal, NP=Nobel
Prize, MPM=Max Planck Medal, DM=Dirac Medal, VNM=John Von
Neumann Medal). Important prizes are rarely awarded soon after
a discovery is made and this bias is well visible in our table.
To overcome this, we add an additional distinguishing criterion
for prize-free papers: if they are described as pioneering works
in a certain domain on Wikipedia, we mark them with $+$. The
last two columns show the paper's ranking given by the Page Rank
score when $\alpha=0.5$ (PR) and the citation count (CR). Bold
labels correspond to the papers not detectable as outliers in
Fig.~\ref{fig:PR_indeg}.}
\label{tab:seminal}
\end{table*}

Outliers in the $(P_x,I_x)$ plane often do not have particularly
high citation counts. When we apply the classical PageRank
algorithm to our data as in~\cite{Chen07}, we observe than many
of them do not receive high PageRank values. The differences
stem, of course, from differences between the algorithms. 
While PageRank is a \emph{reputation} metric~\cite{JoIsBo07}
awarding papers cited by other reputable papers, our approach
focuses on the progeny created by each individual paper. In
consequence, even a paper which is not directly cited by popular
papers can score high if it establishes a new research direction
or a school of thought. In this sense, our approach evaluates
originality of papers. On the other hand, interdisciplinary
works necessarily focus the flow of influence less and hence
they are not likely to score high with respect to the $I_x/P_x$
criterion.

\begin{figure}
\center
\includegraphics[scale=0.28]{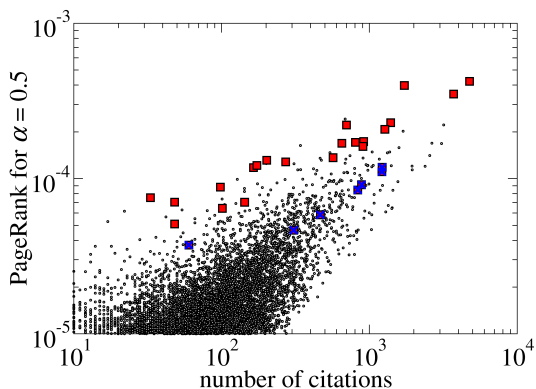}
\caption{PageRank with $\alpha=0.5$ vs citation count (with an
older version of the APS data, a similar plot was already
presented in \cite{Chen07}). Outliers from
Fig.~\ref{fig:seminal} are marked either with red squares (if
they can be considered as outliers also in this figure) and with
blue crosses (if they are not outliers here---these papers have
their number written in bold in Table~\ref{tab:seminal}).}
\label{fig:PR_indeg}
\end{figure}

We finally note that the definition of the PageRank score
$I^\alpha_x$ in Eq.~\ref{Ixa} allows for a meaningful research
of outliers in the $(I^\alpha_x,k^{in}_x)$ plane (see
\cite{Chen07}), similarly as we do in the $(I_x,P_x)$ plane for
the aggregate impact $I_x$. While some papers appear as outliers
in both planes, there are some significant differences which
further demonstrate the distinction between our evaluation
metric and the PageRank (see Fig.~\ref{fig:PR_indeg}). These
differences, marked with bold letters in
Table~\ref{tab:seminal}, correspond to relatively recent but
seminal papers, suggesting that our method is more effective in
removing the inherent time bias of citation data
discussed above.

After showing that our concept of influence quantified by the
$\mathsf{G}$ matrix has its merit, we use it to evaluate
similarity of papers. The basic idea is that papers $x$ and $y$
are similar if they are influenced by the same works (they have
similar ``genetic'' composition). To evaluate this similarity we
take
\begin{equation}
\label{sim-our}
S^*(x,y)=\sum_i \sqrt{G_{ix}G_{iy}}.
\end{equation}
It is also possible to base the similarity on
$\min\{G_{ix},G_{iy}\}$ or $G_{ix}G_{iy}$, for example---we
present here the choice performing best in our numerical tests.
Note that this similarity is not normalized: its lower bound is
zero but the upper bound is bounded only by
$\mathcal{A}_x\cap\mathcal{A}_y$. We stress that $S^*$ is
parameter-free and hence practical to use.

The standard way to evaluate a similarity metric is to test how
well it is able to reproduce missing links in a
network~\cite{Liben-Novell07,TaoLin}. In practice this means
that small part of links (usually $10\%$) is removed from the
network and one attempts to guess the removed links by seeing
which similar nodes are not connected. A similarity metric
which is able to ``repair'' well the network presumably captures
well the network's structure and one may use it also for
other purposes than link prediction. In the case of our
similarity metric $S^*$, we adopt a slightly different approach:
we test how good recommendations it is able to provide to
selected individuals. This change is motivated by potential
practical use of such recommendations for scientists who often
face the problem of searching for relevant literature in their
research field~\cite{McNee2002}. 

Our tests are done as follows. We first divide the data in two
parts: papers published until year 2003 (the sample set---it
contains approximately $75\%$ of all papers) and those published
after 2003 (the probe set). Then we find 20 most-cited articles
published in each core APS journal in 2003 (we consider seven
journals: Phys. Rev. Lett., Rev. Mod. Phys. and Phys. Rev. A--E)
and take their last authors if they published at least one paper
with the APS after 2003. Recommendations are made for each test
author separately on the basis of papers published by this
author in 2003. Denoting the set of papers published by author
$\alpha$ in 2003 as $\mathcal{U}_{\alpha}$, the recommendation
score of paper $x$ is given by its similarity with all $y$ in
this set
\begin{equation}
\label{rec_score}
r_x=\sum_{y\in\mathcal{U}_{\alpha}} S^*(x,y).
\end{equation}
Papers that haven't been cited by author $\alpha$ until 2003
are then sorted according to their score in a descending order
and those at the top represent \emph{personalized
recommendation} for this author.

Resulting recommendations are evaluated using the probe set
which allows us to label as ``relevant'' those papers that were
eventually cited by a given author after 2003. To curb the level
of noise in the results, we discard authors with less than $10$
relevant papers to be guessed. Then we are left with the final
set of $99$ test authors who have on average $116$ relevant
items to be guessed out of almost $340\,000$ papers published
until 2003. To assess the recommendations, we use metrics often
used in the field of recommender systems~\cite{AdoTuzi05}: (i)
precision $P_{100}$ (the fraction of the top 100 places of the
recommendation list occupied by the relevant papers), (ii)
recall $R_{100}$ (the fraction of the relevant papers appearing
at the top 100 places of the recommendation list), (iii) the
average ranking of the relevant papers $q_R$ (expressed as a
fraction of all potentially relevant papers), and (iv) the
fraction of the relevant papers with non-zero score $f_R$. A
good recommendation list should have relevant papers at the top,
\emph{i.e.}, high $P_{100}$ and $R_{100}$ and low $q_R$, and it
should assign non-zero scores to most relevant papers,
\emph{e.g.} high $f_R$ (all these quantities lie in the range
$[0,1]$).

\begin{figure}
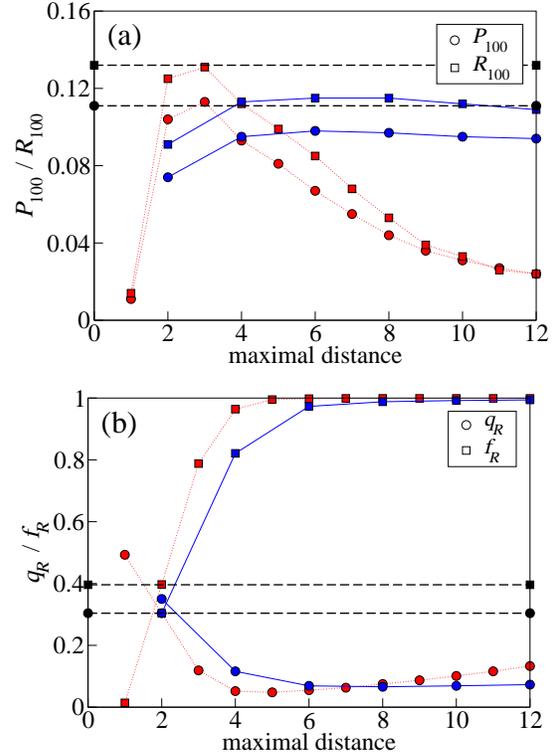

\centering
\vspace*{2pt}
\includegraphics[scale=0.28]{4a}\\[8pt]
\includegraphics[scale=0.28]{4b}
\caption{Precision and recall (a) and average ranking of
relevant items and fraction of ranked relevant items (b) for
$S^{KA}$ (red symbols, dotted lines), $S^*$ (blue symbols,
solid lines) and $S^{RA}$ (black symbols, dashed lines).
$S^{KA}$ shows a strong dependency on the maximal distance with
best $P_{100}$ and $R_{100}$ achieved when the maximal distance
is $3$. However, $q_R$ is only $0.79$ at this point which means
that at this level of truncation, it represents a transition
between local and global similarity metrics. When all powers of
$\mathsf{A}$ are included, $S^{KA}$ performs poorly with respect
to all measured characteristics but $f_R$. By contrast, the
performance of $S^*$ decreases only slightly when the maximal
distance is above eight.}
\label{fig:test}
\end{figure}

To test our similarity, we compare its performance in a
recommendation process with other similarity metrics. Based
on results presented in~\cite{TaoLin}, we have selected three
highly performing metrics: the Common Neighbors similarity (CN),
the Resource Allocation Index (RA), and the Katz-based
similarity (KA). Since they are all defined on undirected
networks, we evaluate them assuming that all links in our data
are undirected. CN simply counts the number of common neighbors
for a pair of nodes. RA does the same but it values less common
neighbors with many connections,
\begin{equation}
\label{sim-RA}
S^{RA}(x,y)=\sum_{z\in\Gamma(x)\cap\Gamma(y)}
\abs{\Gamma(z)}^{-1}
\end{equation}
where $\Gamma(x)$ is the set of direct neighbors of node $x$.
We finally employ a commonly used similarity, KA, which counts
the number of paths between two given nodes with individual
paths weighted exponentially less according to their length
(this similarity has a close relation with the Katz centrality
measure~\cite{Katz}). Denoting the network's adjacency matrix
with $\mathsf{A}$, KA can be written in the form of a series
\begin{equation}
\label{sim-KA}
S^{KA}(x,y)=\sum_{i=1}^{\infty}\beta^i(\textsf{A}^i)_{xy}.
\end{equation}
In our case, we use $\beta=0.75$ which yields slightly superior
performance. \emph{Local} similarities $S^{CN}$ and $S^{RA}$ are
computationally considerably less demanding than \emph{global}
(based on the whole network) similarities $S^*$ and $S^{KA}$.
For practical reasons, we limit the computation of $S^*$ to
papers that are not more than six steps from both $x$ and $y$.
For $S^{KA}$, we limit its summation to the order
$\mathsf{A}^{12}$ (see Fig.~\ref{fig:test} for how these
restrictions affect the results).

Similarities described above can be substituted for $S^*(x,y)$
in Eq.~(\ref{rec_score}), leading to recommendations which can
be in turn compared with those obtained with $S^*$. Test results
can be found in Fig.~\ref{fig:test} where we plot performances
of different algorithms vs the maximal distance used to compute
global similarities. Results for the Resource Allocation Index
are indicated with flat lines while results for the Common
Neighbor similarity are omitted because they are always worse
than for $RA$. In general we see a good performance of $S^{RA}$
with respect to precision and recall. This is because local
metrics rank only a small set of papers (local neighborhoods)
where there is high probability of finding relevant papers. The
drawback is that only a minor part of relevant papers is found
($f_R\approx 0.4$) and their average ranking is poor
($q_R\approx 0.3$).

At the same time, global metrics $S^*$ and $S^{KA}$ are able to
rank almost all relevant objects and achieve much lower average
ranking, but they pay for this enhanced 'variety' with worse
performance at top places of their recommendation lists. When
the maximal distance of five or more is considered (which is
necessary for making $S^{KA}$ a truly global similarity metric
with $f_R\approx1$, $S^*$ significantly outperforms $S^{KA}$
and, from the point of view of recommendation, provides a good
compromise between global and local metrics. This is despite the
fact that $S^{KA}$ and $S^{RA}$ are computed on undirected data
which gives them access to more information: they assign
similarity also to nodes with overlapping progeny, not only to
those with overlapping ancestors as $S^*$ does. Further tests
show that if we prevent $S^{RA}(x,y)$ from accessing this
information, its precision and recall decrease to $0.104$ and
$0.124$ respectively which is comparable to the results obtained
with $S^*$. We may conclude that $S^*$ is a reliable similarity
metric which is able to compete with other known metrics.

In conclusion, our results unveil the value of the passage
probability in random walks on DAGs. On the example of
scientific citations we showed that it allows us to quantify the
influence of a given paper (node) on the others, to identify
seminal and innovative papers (\emph{i.e.}, instrumental nodes
of the network), and to introduce a similarity metric whose
performance is comparable with that of other state-of-the-art
metrics. In this Letter, we aimed at simplicity and hence we
didn't consider additional effects that may have impact on the
interpretation of the analyzed citation data. For example, we
didn't consider that every paper relies on general knowledge
which is however never cited. To reflect that, one could for
example add an artificial node referred by every other node in
the network and repeat the same analysis as we did. Further,
similarly as for PageRank~\cite{WXYM07}, our framework also
lends itself to generalizations based on assigning past
citations with lower weights to better reflect current relevance
or, more generally, trends. We believe that our framework might
prove useful well beyond citation networks as it opens
possibilities for the investigation of asymmetric interactions
in DAGs by exploiting their intrinsic acyclic nature. The
presented ideas and tools can be readily applied
to citation networks related to any kind of intellectual
production such as patents and legal cases. Similar networks of
dependency relations can also be found in biology (phylogenetic
networks and food webs, for example) as well as in other systems
that can be mapped into a DAG, where individuation of
fundamental nodes and estimation of dependency relations within
the graph can be useful and non-trivial tasks.

\acknowledgments
This work was supported by the EU FET-Open Grant 231200 (project
QLectives), by the Swiss National Science Foundation (Grant No.
200020-132253), by the National Natural Science Foundation of
China (Grant Nos. 60973069, 90924011) and the Sichuan Provincial
Science and Technology Department (Grant No. 2010HH0002). We are
grateful to the APS for providing us the dataset.

\end{document}